# Simulation-based assessment of a Bayesian survival model with flexible baseline hazard and time-dependent effects

Iain R. Timmins[1]*, Fatemeh Torabi[2-4], Christopher H. Jackson[4], Paul C. Lambert[,5,6], Michael J. Sweeting[1,7]

[1]Statistical Innovation, Oncology R&D, AstraZeneca, Cambridge, UK

[2]Institute of Continuing Education, University of Cambridge, Cambridge, UK

[3]Dementias Platform UK, Swansea University, Swansea, Wales, UK

[4]MRC Biostatistics Unit, University of Cambridge, Cambridge, UK

[5]Cancer Registry of Norway, Norwegian Institute of Public Health, Oslo, Norway

[6]Department of Medical Epidemiology and Biostatistics, Karolinska Institutet, Sweden

[7]Department of Population Health Sciences, University of Leicester, UK

*email: iain.timmins@astrazeneca.com

# Abstract

There is increasing interest in flexible parametric models for the analysis of time-to-event data, yet Bayesian approaches that offer incorporation of prior knowledge remain underused. A flexible Bayesian parametric model has recently been proposed that uses M-splines to model the hazard function. We conducted a simulation study to assess the statistical performance of this model, which is implemented in the `survextrap` R package. Our simulation uses data generating mechanisms of realistic survival data based on two oncology clinical trials. Statistical performance is compared across a range of flexible models, varying the M-spline specification, smoothing procedure, priors, and other computational settings. We demonstrate good performance across realistic scenarios, including good fit of complex baseline hazard functions and time-dependent covariate effects. This work helps inform key considerations to guide model selection, as well as identifying appropriate default model settings in the software that should perform well in a broad range of applications.

# Introduction

Statistical analysis of time-to-event data arises across many fields of applied research and is especially common in medical settings such as clinical trials and epidemiological studies. In many applications it is important to smoothly model the baseline hazard function as well as understand the effect of covariates on survival time, enabling predictions of both absolute and relative risk. Often, however, the distribution of survival times and their dependence on covariates cannot be adequately captured by standard parametric models (such as the exponential, Weibull and Gompertz models, among others), and so the use of flexible parametric models has become increasing common. This includes the Royston-Parmar approach of using a restricted cubic spline on the log cumulative baseline hazard [1], as well as other methods such as fractional polynomials [2] and piecewise models [3]. Alternative approaches based on semi-parametric and non-parametric models can also be used [4-6]. Many of these methods have widely used implementations in standard software packages [7, 8].

While flexible modelling approaches have been implemented from a frequentist perspective, Bayesian methods have received less research attention and are more scarcely used in practise. This likely owes to additional complexities related to prior specification, model comparison, computation burden, and fewer existing user-friendly implementations [9, 10]. A Bayesian modelling perspective has advantages, enabling the explicit use of external data and evidence sources to be expressed quantitively through prior beliefs, while such methods also lend themselves naturally to being extended to incorporate multiple sources of data through hierarchical models and evidence synthesis frameworks. In health and medical contexts, this has led to Bayesian survival approaches being used for applications such as the design of efficient clinical trials that incorporate historic data sources [11], as well as their use in Health Technology Assessment [12]. One important potential application of the `survextrap` R package has been in providing improved long-term projections of patient survival from clinical trial data, which are used in the context of health economic models (hence the package name relates to the term "survival extrapolation"), though we do not explore this more specific application of the software in this work.

In this study we assess the capability of the `survextrap` package for generic survival modelling, rather than for extrapolation, as we envisage this Bayesian modelling approach to have broader potential uses across other fields. This is the first comprehensive evaluation of the performance of the `survextrap` Bayesian model across a variety of settings.

The `survextrap` Bayesian survival model has many attractive features and potential advantages over standard parametric models and other approaches, such as flexible modelling capabilities, and the possibility of supplementing observed data with prior information. Nevertheless, there are many modelling choices that need to be considered. To ensure best use of `survextrap`, we performed simulation investigations in application to oncology clinical trials to compare a range of flexible models, varying the M-spline specification, smoothing procedure and prior specification, along with other statistical and computational modelling choices that can be chosen by the user. This helps provide recommendations for model settings that should perform well across a broad array of applications, as well as helping to identify the best default settings within the software (as many users are likely to depend on these). We assessed the performance of different model choices via standard frequentist measures such as bias and coverage. This work helps guide model specification and improve user-confidence in implementing a flexible Bayesian modelling approach to survival data.

# Methods

## Bayesian survival model

The `survextrap` R package fits a flexible parametric Bayesian survival model using M-splines to model how the hazard function changes over time. The hazard function is defined as a weighted sum of $n$ basis functions [16, 17] $b_i(t)$ which take the form:

$$h(t) = \eta \sum_{i=1}^{n} p_i \, b_i(t)$$

with scale parameter $\eta$ and basis coefficients $p_i$ where $\sum_i p_i = 1$. The basis functions are polynomials (cubic by default) and are totally determined by the knots and degree of the M-splines specified. The basis functions are restricted to be positive, so that the resulting hazard is a smooth, positive-valued function.

In the current default, `survextrap` uses $n = 10$ basis functions that are defined on a set of knots placed at quantiles of the uncensored survival times. As the number of knots increase, the hazard becomes more flexible with potentially more turning points. In `survextrap`, the default number of knots is large to accommodate all plausible shapes. However, hierarchical priors protect against the risk of over-fitting and help to smooth the data. This is similar to the use of frequentist penalized likelihood methods that help control the smoothness of the hazard function [18].

Priors are placed on the scale parameter $\eta$ and each of the basis coefficients $p_i$. A hierarchical prior on the $p_i$ is specified by defining $\log(p_i/p_1) = \gamma_i$ with $\gamma_0 = 0$, $\gamma_i = \mu_i + \sigma \epsilon_i$. The prior means $\mu_i$ are fixed and are determined to correspond to a constant hazard, while the parameter $\sigma$ controls the smoothness of the hazard curve, relating to the level of departure from a constant hazard. The random effects $\epsilon_i$ are specified using a weighted random walk with $\epsilon_1 = 0$ and $\epsilon_i \sim \text{Logistic}(\epsilon_{i-1}, w_i)$, where the weights $w_i$ depend on the distance between the knots (see Phillippo et al. [14] for exact definition), or alternatively by using an exchangeable model with $\epsilon_i \sim \text{Logistic}(0,1)$. The default priors are $\eta \sim N(0, 20)$, and $\sigma \sim \text{Gamma}(2,1)$, though the package provides simulation-based procedures to determine priors that match a user's substantive judgement about expected survival and hazard variations.

The Bayesian model can be extended to allow the hazard to depend on explanatory variables; the effect of these can either be constant through time with the assumption of proportional hazards, or time-varying (i.e. a non-proportional hazards model). In the proportional hazards model, the scale parameter is redefined with covariates $\boldsymbol{\beta}$ and explanatory variables $\mathbf{x}$:

$$\eta(\mathbf{x}) = \eta_0 \exp(\boldsymbol{\beta}^T \mathbf{x})$$

In addition, a flexible non-proportional hazards model allows explanatory variables to affect the M-spline coefficients through a multinomial logistic regression where $\gamma_i(\mathbf{x}) = \mu_i + \boldsymbol{\delta}_i^T \mathbf{x} + \sigma \epsilon_i$. The $s$th element of the vector $\boldsymbol{\delta}_i$ describes the amount of departure from a proportional hazards model for the $s$th covariate in the region of time associated with the $i$th basis term. Hence, if $\delta_{is} = 0$ for all $i$, then the $s$th covariate follows proportional hazards. A hierarchical prior is used for these coefficients such that $\delta_{is} \sim N(0, \tau_s)$, which smooths the effects over the time regions. A relatively weak Gamma(2,1) non-informative prior is the default option for each of the $\tau_s$.

A number of additional features exist that the user can specify including incorporating aggregate survival data from external datasets in an evidence synthesis model, the use of cure models, modelling in an excess hazards framework, and sensitivity analyses with the waning of treatment effects in the long-term. These features are not investigated in this simulation study.

## Simulation design and data generation

A simulation study was undertaken for two different trials with distinct hazard functions, where in both cases the true hazard functions were known, allowing assessment of the frequentist properties of Bayesian flexible parametric models using `survextrap`. We simulated survival times from a flexible parametric Royston-Parmar model [1], which was fitted to each of the two trials.

### Data generating mechanisms

For our first case study we considered overall survival for patients treated with radiotherapy plus cetuximab for head and neck cancer from Bonner et al. [19], which will be referred to as the Cetuximab OS case study hereafter. We reconstructed patient-level time-to-event data for the radiotherapy plus cetuximab arm (N = 211) from the Kaplan-Meier curve for overall survival using WebPlotDigitizer (version 4.5) [20] based on the algorithm by Guyot et al. [21]. We found that a Royston-Parmar spline model with 3 knots fitted in `flexsurv` [7] adequately captured the fit of the trial data (Figure 1a).

As a second case study, we also analysed progression free survival for patients treated with nivolumab for advanced melanoma from Larkin et al. [22], which we refer to as the Nivolumab PFS case study, imputing patient level data from the Nivolumab PFS (N = 316) Kaplan-Meier curves as described above. For our true model we used a Royston-Parmar model with 6 knots fitted to Nivolumab PFS data. This represented a more widely-varying hazard function than the one derived from the cetuximab data (Figure 1b).

For these underlying true models, we generated 1,000 single-arm datasets with a trial sample size of N = 200 patients. Simulated survival times were generated using the `simulate` function in `flexsurv` [7]. Administrative censoring was applied at 5 years.

To understand the performance of `survextrap` for evaluating treatment effects, we used the model for radiotherapy plus cetuximab OS described above as our control arm, and further defined an active arm by specifying a hazard ratio across trial follow-up time that is relative to the cetuximab OS control arm. We consider three treatment effect scenarios; a constant treatment effect with proportional hazards (scenario one), an immediate treatment effect followed by treatment waning (scenario two) and a delayed treatment effect followed by treatment waning (scenario three). The closed-form expressions for the hazard ratio functions in each scenario are shown in Supplementary Table 1 and can be seen in Figure 2.

To simulate event times for two arms we implemented the cumulative hazard inversion method [23] in R software. Since the cumulative hazard function for the active arm does not have a closed-form solution for these scenarios, we used numerical integration to derive the cumulative hazard using Gauss-Legendre quadrature with 100 nodes. For this two-arm setting, we generated 1,000 datasets with a trial sample of size of N = 400 patients, with 200 patients in each arm.

**Estimands**

Our primary estimand was the restricted mean survival time (RMST) at 5 years for the single treatment arm setting and the difference in RMST at 5 years (RMSTD) for the treatment effect between active and control. For each simulation setting we generated one very large dataset (N=$10^8$) without censoring and empirically calculated the RMST and RMSTD using their sample mean. We took these to be the true values for each data generating setting, which were accurate to 2 decimal places based on Monte Carlo standard errors of the sample means. Further details of the estimands and performance measures are given in Appendix 1.

**Figure 1:** Data generating mechanisms, survival and hazard plots for the Cetuximab (overall survival) and Nivolumab (progression-free survival) case studies.

**(a) Cetuximab OS**

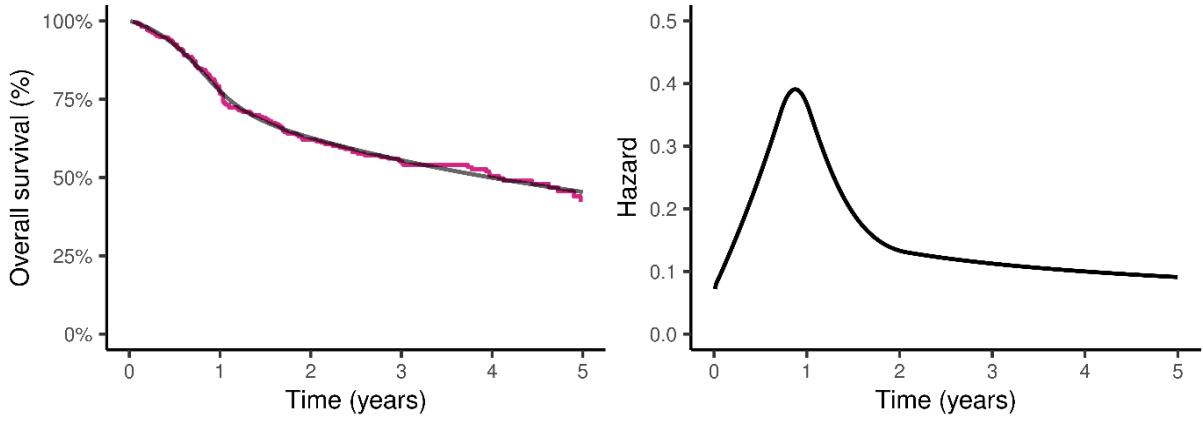

**(b) Nivolumab PFS**

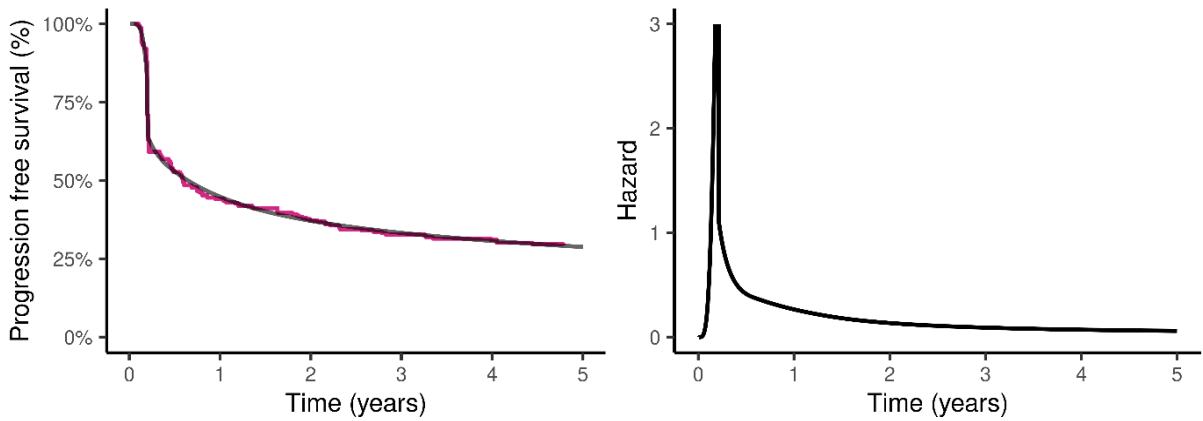

**Figure 2:** Data generating mechanisms, survival and hazard ratio plots for 3 treatment effect scenarios.

**(a) Scenario 1: Constant effect (proportional hazards)**

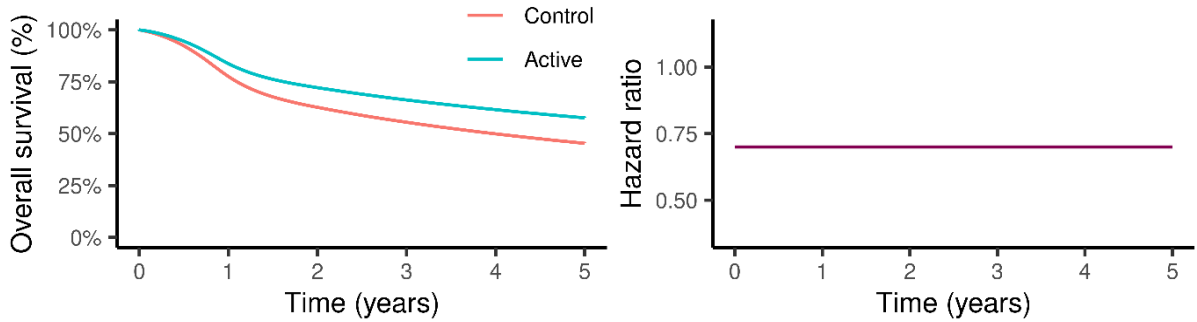

**(b) Scenario 2: Waning effect**

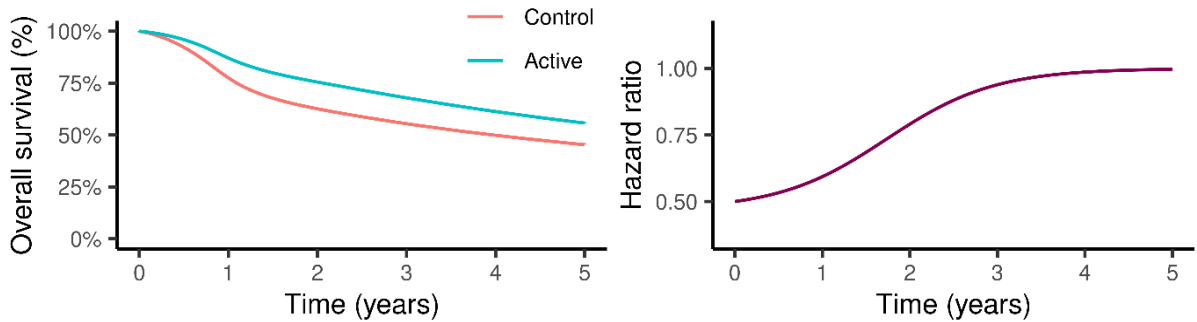

**(c) Scenario 3: Delayed then waning effect**

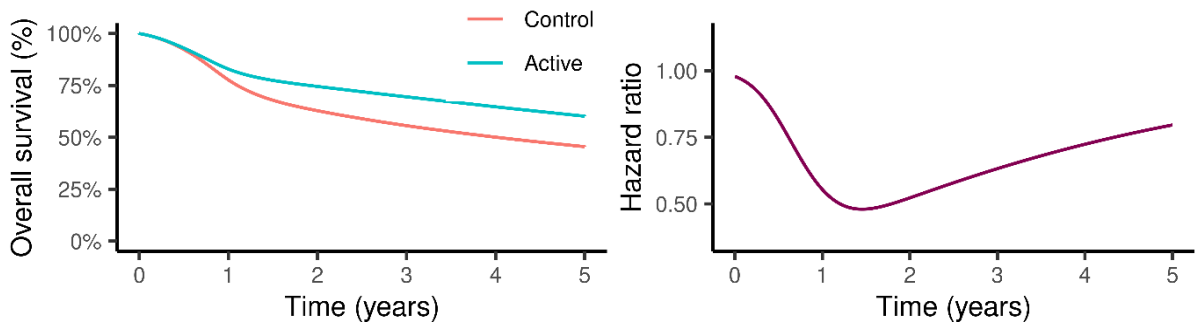

**Analysis of simulated data**

Bayesian flexible parametric models were fitted to each simulated dataset using the `survextrap` package (version 0.8.12) [15]. We considered a range of Bayesian models, with different specifications of the cubic M-spline hazard function, while further varying the priors for the smoothing parameter $\sigma$. We varied the M-spline degrees of freedom (df) between df = 3, 6 and 10, and also varied the $\sigma$ prior between Gamma(2,1), Gamma(2,5) and Gamma(2,20). To interpret these different priors for $\sigma$, we note that a Gamma(2,1) prior allows for a very flexible hazard function and gives an upper 90% credible limit for the ratio between the 90% and 10% quantiles of the hazard $h(t)$ (with respect to time) of $4.2 \times 10^8$, while a Gamma(2,20) prior is much more restrictive and gives an upper 90% credible limit of 1.6 for this ratio. We further varied the choice of prior for the spline coefficients, evaluating models with the weighted random walk and the exchangeable model. This led to 18 combinations of model settings. A further comparison was made of the standard versus smoothed M-spline basis options, details of which are in the Appendix 2.

For each of these 18 Bayesian models, we further varied the choice of Bayesian computation method (from the `rstan` package [24], that `survextrap` calls). Firstly, we fitted all models using the default MCMC sampling routine with 4 chains and 2,000 iterations. We additionally analysed all simulated datasets using the fast method of evaluating the posterior median using a Laplace optimisation approximation to the full posterior distribution.

Additionally, we compared the `survextrap` Bayesian model with frequentist spline-based approaches, where we used Royston-Parmar models implemented in `flexsurv` and `rstpm2` R packages. These models use natural cubic splines on the cumulative hazard scale. We fitted Royston-Parmar models with 2, 3, 4, 6 and 10 df. The standard errors on RMST estimates were computed using the bootstrap method with 1,000 iterations for `flexsurv` and the delta method for `rstpm2`.

For each of the three treatment effect scenarios we assessed the performance of the `survextrap` non-proportional hazards models. For the non-PH models we fitted models with 3, 6 and 10 df, while varying the strength of the non-proportionality smoothness prior $\tau$ between Gamma(2,1), Gamma(2,5) and Gamma(2,20), and keeping the prior for the mean log hazard ratio at a vague Normal(0,20). For comparison, we also fitted the `survextrap` proportional hazards model with standard settings of df = 10, $\sigma \sim$ Gamma(2,1) and random walk prior.

For benchmarking comparisons, we further evaluated treatment effects in both `flexsurv` and `rstpm2`. In each case we modelled the control arm using a baseline hazard with 2, 3, 4, 6 and 10 df. For the time-varying effect in `flexsurv`, treatment covariates were placed on each spline coefficient as ancillary parameters, so the model was fully stratified by treatment, with the time-varying treatment effect having the same form of spline as the baseline hazard (same degree, knot locations and basis functions with different coefficients). In `rstpm2` there is the additionally flexibility with the `tvc` argument to specify time-varying effects with a spline of a different form to the baseline hazard, and we varied the spline for the time-varying treatment effect between 1, 2, 3 and 6 df while varying the baseline hazard df as described above.

# Results

## Estimating the RMST for a single treatment arm

### Cetuximab OS Case Study

For the Cetuximab OS Case Study, the true survival was 62.7%, 55.5% and 45.4% at 2, 3 and 5-year timepoints, respectively, while the true restricted mean survival (RMST) at 5 years was 3.19 years (Figure 1a).

Across 1,000 simulated datasets with 200 patients, the average number of events within 5-year follow-up time was 113 (s.d. = 7). The performance measures for estimating 5-y RMST for cetuximab OS are shown in Figure 3a. For the `survextrap` model using default options of 10 df for the M-spline, a smoothness prior $\sigma \sim$ Gamma(2,1) and a random walk prior on the spline coefficients, the mean estimate of 5-y RMST across 1,000 simulations was 3.19 (Monte Carlo standard error (MCSE) = 0.004), which corresponds to an absolute bias of 0.007 (0.004) years and relative bias 0.23% (0.14%). For this model, the posterior standard deviation of 5-y RMST, averaged across simulations, was 0.134 (0.0001), which matched the true sampling variability, measured by the empirical standard error of 0.136 (0.003). The coverage of the 95% credible intervals for this model was 94.5% (0.7%). The fitted survival and hazard curves were centered around the true curve and had the same shape (Supplementary Figure 1).

We also observed comparable model fit when varying the df of the M-spline hazard function whilst keeping the default smoothing prior of Gamma(2,1) (Figure 3a), with low absolute and relative bias (<1%), and only slight undercoverage. When increasing the strength of the smoothness prior $\sigma$ to Gamma(2,5) and Gamma(2,20), we observed increases in bias and poorer coverage (Figure 3a). We also found that these stronger priors only achieved sufficient shrinkage when using a weighted random walk prior on the spline coefficients (Supplementary Figure 1), while in contrast, there was minimal shrinkage towards a constant hazard with the exchangeable model (Supplementary Figure 2). Nevertheless, the exchangeable model gave comparable performance to the random-walk model in terms of bias across the scenarios investigated (Supplementary Figure 3).

We further assessed the MCMC convergence performance of the `survextrap` models (Supplementary Figure 4). With df = 10, $\sigma \sim$ Gamma(2,1) and a random walk prior, we found that 0.1% of simulation replicates had poor convergence as indicated by $\hat{R} > 1.05$. We observed that 29% of the 1,000 simulation replicates reported at least one divergent transition in the MCMC sampler, and among the chains with at least one divergent transition, the proportion of the total iterations that diverged remained low at 0.5%. As such, any divergences in the MCMC sampler were minor and did not appear to adversely impact estimation performance. For this model we also found reasonable MCMC sampling efficiency across the posterior, with only 1.7% of simulation replicates having low bulk effective sample size (<400, where the combined chains had length 8,000). Convergence was also improved using the random walk prior compared to the exchangeable prior (Supplementary Figure 5).

**Nivolumab PFS Case Study**

For the Nivolumab PFS case study the true 5-y RMST was 1.98 years (Figure 1b), and the true PFS survival probability at 5 years was 28.8%. A `survextrap` model with default settings (df = 10, $\sigma \sim$ Gamma(2,1) and a random walk prior) gave a mean estimate of 5-y RMST of 1.97 (0.005) years, with minimal absolute bias of -0.006 (0.005) years and relative bias -0.29% (0.24%), and reasonable coverage probability of 94.2% (0.8%) (Figure 3b). This model had sufficient flexibility to adequately capture the sharp hazard function and steep gradient of the true survival curve (Supplementary Figure 6).

When varying the model parameters, we noted that for a model with an M-spline hazard function with 3 df, there was marked downward bias in estimates of 5-y RMST (Figure 3b), as this model did not capture the sharp gradient present in the true survival curve (Supplementary Figures 6 and 7).

**Fast Laplace approximation**

We assessed the Laplace approximation method within `survextrap` as a fast approximation to full MCMC sampling, with results presented in Supplementary Figure 8. The runtime for the Laplace optimisation approximation was ~ 0.4 seconds per model fit compared to ~50 seconds using full MCMC, for models with 10 df (Supplementary Table 2). All models fitted using the optimisation approximation converged. For all models tested and for both case studies, we observed that the optimisation approximation had point estimates comparable to the full MCMC sampler, with similar bias. However, the model posterior standard deviations were higher than for models estimated using full MCMC, resulting in over coverage of 97.0% (0.5%) under the default settings (df = 10, $\sigma \sim$ Gamma(2,1)).

**Comparison with standard frequentist models**

Performance measures for estimates of 5-y RMST using Royston-Parmar spline approaches, implemented in both `flexsurv` and `rstpm2` R packages, are shown in Supplementary Figure 9. In comparison to the `survextrap` model with default settings, we found that Royston-Parmar models with sufficient flexibility had comparable performance when estimating 5-y RMST. For the Cetuximab OS case study, we noted that Royston-Parmar models with at least 3 df performed well, with very low bias (relative bias <0.9%), whilst for the Nivolumab PFS case study the best performance was achieved for models with at least 4 df (relative bias <1.1%). We noted that Royston-Parmar models with df = 2, equivalent to a standard parametric Weibull model, had relative bias of 3.8% (0.13%) and 6.7% (0.25%) for the Cetuximab OS and Nivolumab PFS case studies, respectively, highlighting the dangers with using an under-parameterised parametric model. For the models with sufficient df, the model standard errors were well-calibrated.

**Figure 3:** Performance measures for the posterior median RMST at 5-y based on simulated data from the Cetuximab OS and Nivolumab PFS case studies using a random walk prior for `survextrap`.

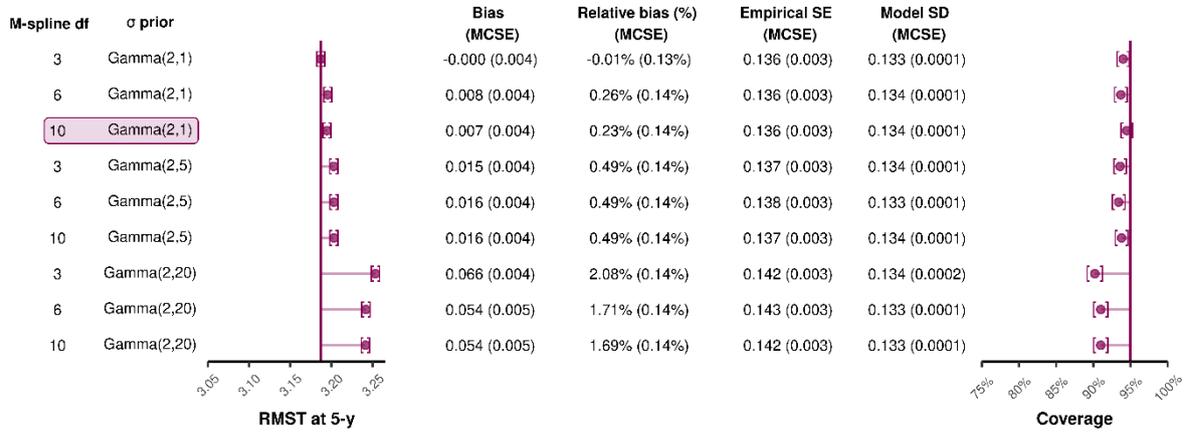

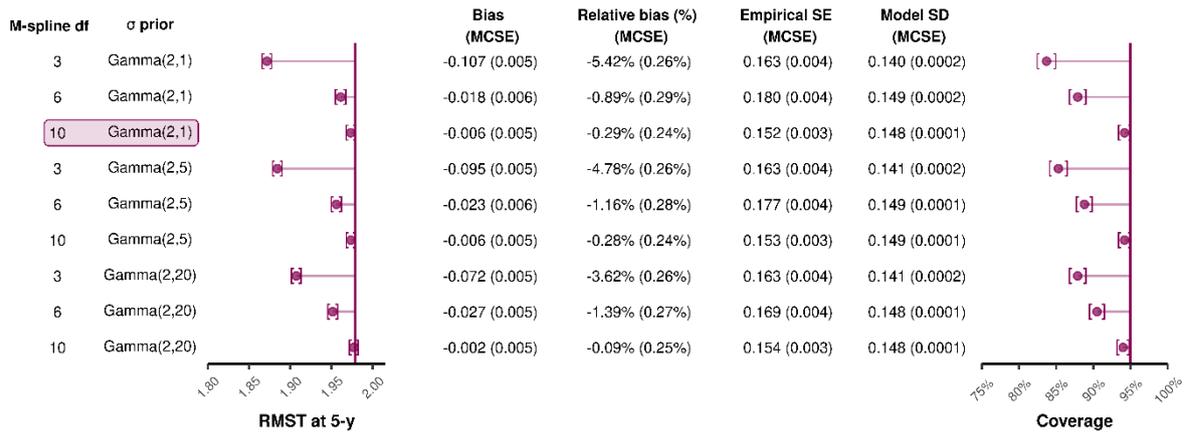

Highlighted rows show the default settings in `survextrap`.

**Estimating the difference in RMST between two treatment arms**

For the case study with two treatment arms the true underlying survival curves and treatment effects for each of the three treatment effect scenarios are shown in Figure 3. The true difference in restricted mean survival time (RMSTD) at 5 years for scenario one (constant treatment effect, proportional hazards), two (waning treatment effect) and three (delayed treatment effect followed by waning) was 0.44, 0.51 and 0.53 years, respectively. The performance measures for estimating the difference in RMST at 5 years are shown in Figure 4, where all models have a random walk prior on the spline coefficients and with a Gamma(2,1) prior for $\sigma$.

For scenario one with constant treatment effect, where the true underlying treatment effect was constant with hazard ratio (HR) = 0.7, we found that all `survextrap` models estimated this treatment effect reasonably well. For the `survextrap` proportional hazards (PH) model (df = 10) the mean 5-y RMSTD was 0.43 (0.006) years, corresponding to a bias of -0.006 (0.005) years and relative bias -1.32% (1.20%). The `survextrap` non-PH model with df = 10 and non-proportionality smoothing prior $\tau$ ~ Gamma(2,1) showed slight underestimation of the treatment effect with bias of -0.011 (0.006) years and relative bias -2.59% (1.26%). For the `survextrap` non-PH models, we found that strengthening the non-proportionality prior on $\tau$ to Gamma(2,5) and Gamma(2,20) lead to similar estimates to the PH model in this scenario (Figure 4a), as expected.

For scenarios two (waning treatment effect) and three (delayed treatment effect followed by waning), where the underlying true models had time-varying treatment effects (Figure 3b and 3c), we observed improved estimation performance for `survextrap` non-PH models with fewer df (df = 3 or 6) compared with the default setting of df = 10 (Figure 4b and 4c). For scenario two (waning treatment effect), the `survextrap` non-PH model with df = 10, $\tau \sim$ Gamma(2,1) had more notable bias of -0.035 (0.006) years and relative bias -6.87% (1.10%), whilst under a more simplified non-PH model (df = 3, $\tau \sim$ Gamma(2,1)), the absolute bias reduced to -0.028 (0.006) years with relative bias of -5.54% (1.11%), though bias was still evident. A similar pattern was found for scenario three (delayed treatment effect followed by waning), with the non-PH model with df = 10, $\tau \sim$ Gamma(2,1) having a relative bias of -2.90% (1.08%), while for a less complex model (df = 3, $\tau \sim$ Gamma(2,1)), the bias remained but was slightly reduced -2.65% (1.08%).

The visual fit of the estimated hazard ratio for a random sample of 50 simulation replicates is shown in Figure 5. We note that the non-PH models with a Gamma(2,1) prior for $\tau$ show reasonable closeness of fit to the underlying true hazard ratio, though power is perhaps limited at this trial sample size. We note that strengthening the non-proportionality prior $\tau$ from Gamma(2,1) to Gamma(2,5) flattened out the estimated time-varying treatment effects, leading to underestimation of the 5-y RMSTD in Scenarios 2 and 3 (Figure 4), suggesting this flattening was excessive. Additionally, when fitting models with an exchangeable prior on the spline coefficients (Supplementary Figures 10 and 11), we observed a higher number of turning points and broad divergences from the true treatment effect over the trial period suggestive of overfitting. Model runtime was higher for the non-proportional hazards models and models with higher df (Supplementary Table 3).

**Comparison with standard frequentist models**

Performance measures for treatment effects using frequentist Royston-Parmar spline approaches implemented in `flexsurv` and `rstpm2` are shown in Supplementary Figure 13. For scenario one (constant treatment effect), models with 2 df (equivalent to a Weibull standard parametric model) displayed slight bias, while all other models demonstrated unbiased estimation (within MCSE) of 5-y RMSTD. For scenario two (waning treatment effect), we found only slight bias for all models with at least 3 df on the time-varying effect and at least 4 df on the baseline spline, which reflected improved performance over the `survextrap` models. Similarly, in scenario three (delayed treatment effect followed by waning), we noted that unbiased estimates (within MCSE) were achieved for more flexible models with at least 4 df for both the baseline hazard and time-varying effect. Model and empirical standard errors were closely aligned, leading to good coverage at the 95% level for the models with sufficient df.

**Figure 4:** Performance measures for the difference in RMST at 5-y based on simulated data from the Cetuximab OS case study for the control arm and under three scenarios for the time-varying treatment effect, investigating fitted `survextrap` models that use a random walk prior and a default Gamma(2,1) prior for $\sigma$ and where the df and prior for $\tau$ are varied.

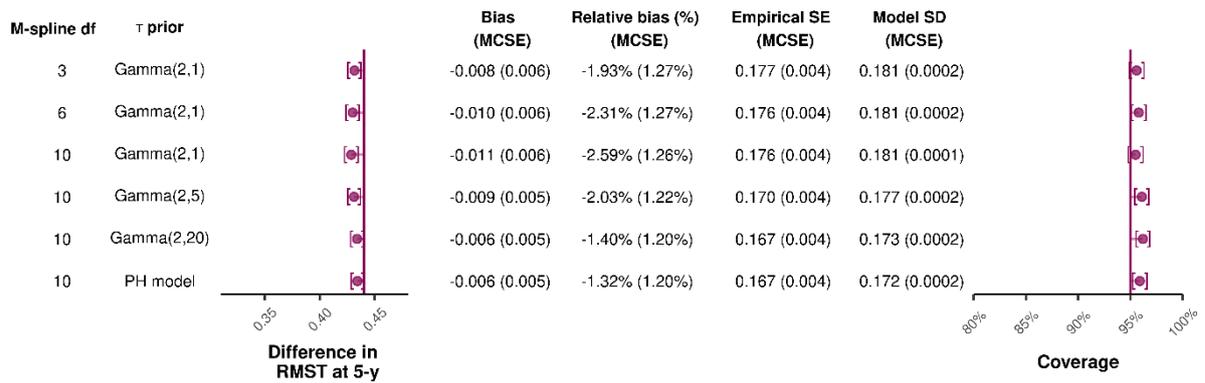

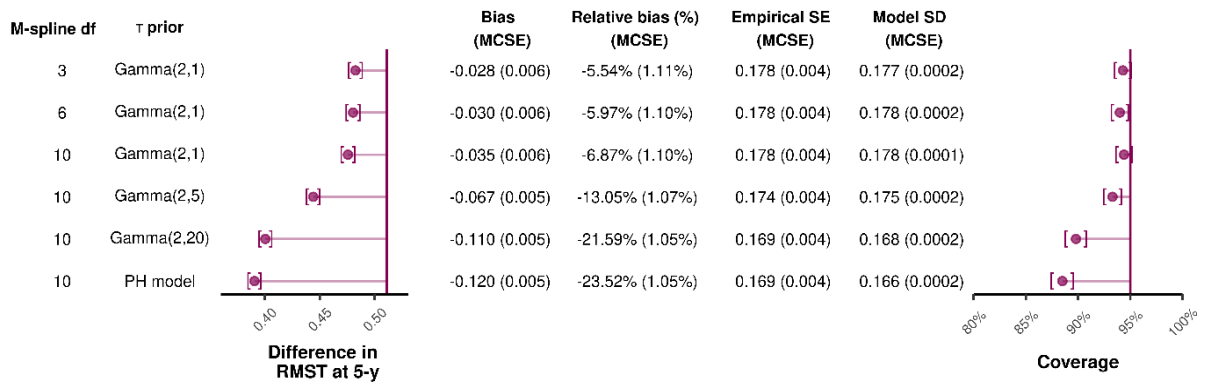

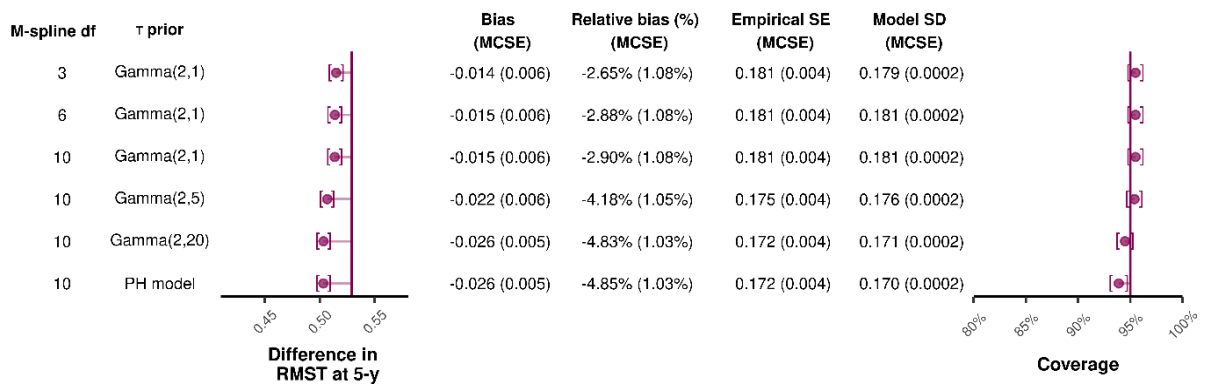

**Figure 5:** Hazard ratio plots for a non-proportional hazards `survextrap` model fitted using a random walk prior and varying the df and non-proportionality smoothness prior $\tau$, based on 50 simulated datasets under three treatment-effect scenarios.

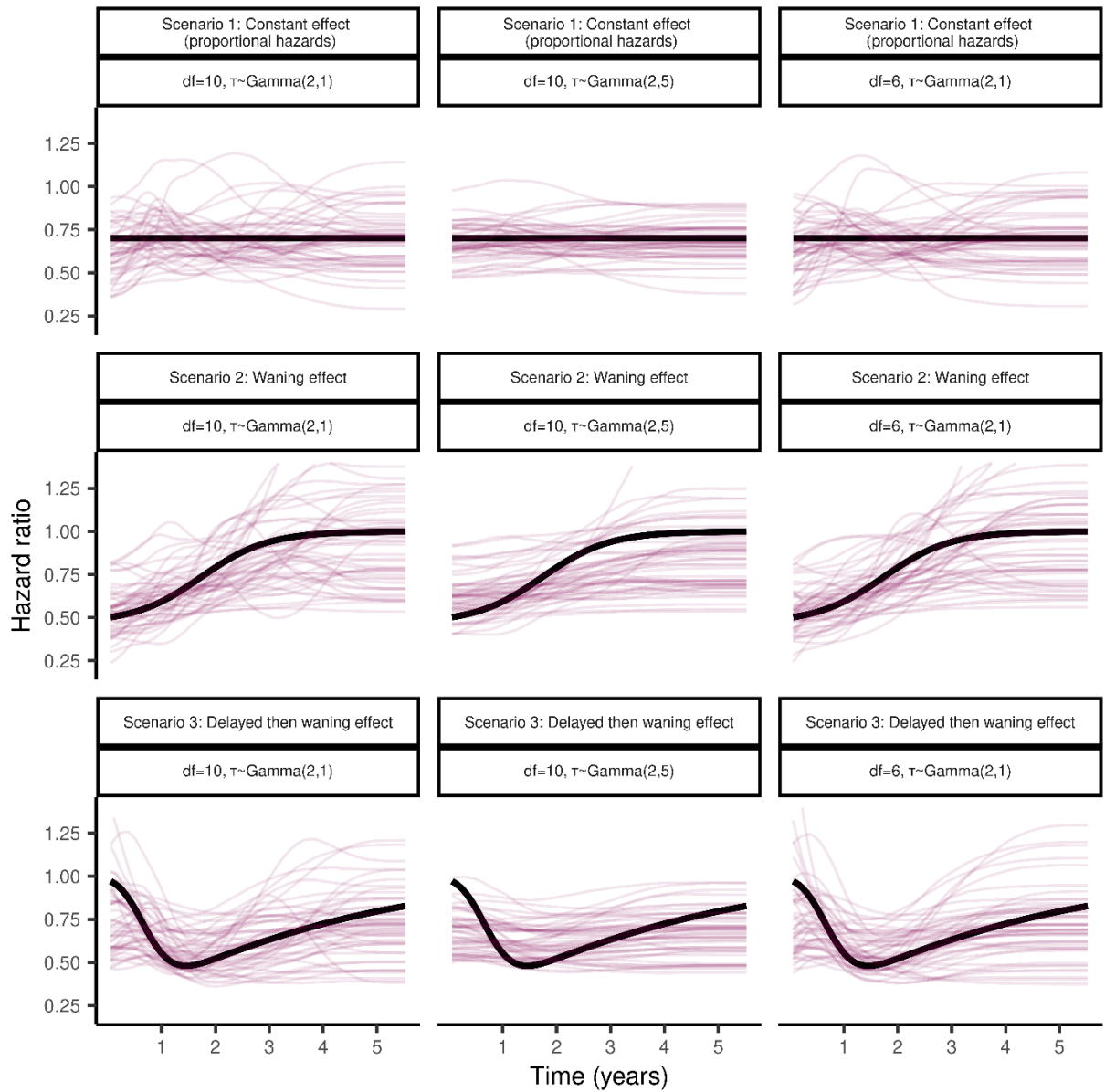

# Discussion

In this study, we have sought to understand the performance of the `survextrap` R package for flexible Bayesian modelling of survival data through a simulation approach applied to case examples from oncology clinical trials.

A primary aim was to help provide users with an understanding of which model specifications best capture the survivor functions through the presentation of the bias and coverage of the estimated RMST at 5-years, and the difference in RMST between two treatment arms. We compared a range of models with different choices for the flexibility of the baseline hazard function, which is dictated both by the number of knots in the M-spline and the prior distributions for the coefficients of the spline. As a starting point for model selection, we identified that a model with 10 degrees of freedom, a smoothness prior of $\sigma \sim$ Gamma(2,1), and a weighted random walk prior on the spline coefficients, demonstrated consistently good performance across all scenarios considered. Specifically, for estimating mean survival (5-y RMST) in a single arm, the relative bias was very low (<1%) and the flexibility in the M-spline was sufficient to capture sharply varying hazard functions without evidence of overfitting, while achieving good computation convergence of the MCMC chains. For estimating the difference in mean survival times across two arms, the relative bias was also quite low at ~5% across scenarios. This model specification should provide a useful set of default settings for `survextrap`, as some users may need to rely on these. Though we suggest that users should check the fit of their model to their own data, and consider whether fit might be improved with other choices. Alternative modelling choices, can be assessed formally using a cross-validatory procedure [25].

As part of this investigation we evaluated the performance of a weighted random walk prior on the M-spline coefficients, which is a modelling approach proposed recently by Phillippo et al. [14] adapted from Li and Cao [26], that has been newly implemented in `survextrap`. We found the weighted random walk prior achieved effective smoothness of the hazard, while in contrast, using an exchangeable random effect prior did not provide sufficient shrinkage toward a constant hazard, consistent with previous findings [14].

We also considered the choice of computational method that `survextrap` uses for implementing Bayesian estimation. The default Markov Chain Monte Carlo procedure based on Hamiltonian Monte Carlo (HMC, from the Stan software [27]) can be computationally demanding, in particular for larger datasets and models with many predictors. The package provides (also via Stan) some fast approximate Bayesian inference procedures, but the accuracy of these approximations has not been thoroughly assessed for these models. We compared the accuracy of estimates from the full MCMC procedure with a fast Laplace approximation based on determining the posterior mode by optimisation. While the fast approximation method gave good point estimates, uncertainties were not as well calibrated as the MCMC approach, especially when model flexibility was poorly specified. We also found the HMC sampler had good convergence properties in realistic trial datasets. A pragmatic recommendation could be to use the Laplace approximation method for model exploration, considering goodness of fit and leave-one-out cross-validation (LOOCV) [25], and then applying the more intensive full MCMC for final analysis.

We also provided a detailed analysis of time-dependent effects, considering a range of scenarios inspired from oncology trials. We found non-proportional hazards (non-PH) models gave broadly good estimation, even if the underlying truth was proportional hazards, though there was some evidence of poorer fit under a treatment effect waning scenario. We noted however that frequentist Royston-Parmar models (natural cubic splines) did outperform `survextrap` in some non-PH settings. The best `survextrap` model tended to depend slightly on the setting, as we found potentially fewer degrees of freedom were required for modelling treatment effects. We suggest that model selection should be performed that considers a range of models with different priors, flexibilities, including both PH and non-PH models.

We have not comprehensively investigated all modelling choices, such as the knot locations, which were left data-driven based on event quantiles, along with the degree of M-spline, which was set as cubic throughout. Nonetheless, previous studies [28, 29] using M-spline basis functions for survival modelling and other applications have shown robustness to these specifications. Note also that the Bayesian models investigated in this paper used the package's default, weakly informative priors. If strong priors are used that conflict with the truth, these may result in bias. In keeping with the design of `survextrap`, we suggest that further work should evaluate the modelling performance for longer-term projections of survival, especially considering the inclusion of prior beliefs from clinical expert opinion or historical trial data, as well as the impact of incorporating real-world evidence sources.

In summary, our simulation results demonstrate the `survextrap` flexible modelling approach has good performance across a range of realistic scenarios, correctly identifying complex baseline hazards and time-dependent covariate effects. Our work helps further provide users with an understanding of the key modelling choices to help guide the process of model selection, ensuring best performance is achieved.

## Competing Interests

## Acknowledgments

IRT is a fellow of the AstraZeneca Postdoctoral Research Programme. CHJ is funded by the Medical Research Council, programme number MC_UU_00040/4. FT is funded by the UKRI-MRC - programme number MR/T033371/1.

# Supplementary Information: Simulation-based assessment of a Bayesian survival model with flexible baseline hazard and time-dependent effects

# Supplementary Information - Contents

## Appendix 1. Estimands and performance measures

We evaluated the frequentist properties of the different models, considering the bias, empirical standard errors and model-based posterior standard deviation, and coverage, while reporting the Monte Carlo standard errors for each (see Morris et al. [1] for full definitions of these performance measures). These measures were computed using the rsimsum R package [2]. For the single-arm cases we compared estimates to the true values obtained from the underlying parametric model derived in flexsurv. We further plotted the fitted survival and hazard function for a random sample of 50 simulation replicates, to evaluate the similarity of their shapes (over time) to the true hazard and survival.

For the treatment-effect, we evaluated the difference in RMST at 5-y (RMSTD) as our estimand. Since the true value cannot be computed using a closed-form expression we instead evaluated the sample estimate of 5-y RMSTD from a very large simulated data set (N = $10^8$) and took this to be the truth (accurate to 2 decimal places). We also further visually assessed the hazard ratio estimates from the survextrap non-PH models against the true values.

Finally, for the models fitted using the full MCMC sampler we assessed how the choice of Bayesian model impacted computational stability using a range of diagnostics. For each model, we first evaluated the proportion of the 1,000 simulations for which "divergent transitions", arising from problems with the Hamiltonian Monte Carlo procedure, were reported by rstan, or any other rstan warning message. We further assessed whether the models displayed high $\hat{R}$ values (>1.05), which are suggestive of poor mixing or convergence of the MCMC procedure. We also identified models with low effective sample size (bulk or tail ESS <400), which measure sampler efficiency and reliability of the Monte Carlo estimates of the posterior quantiles from the sample. For models fitted using the Laplace optimisation approximation method, we computed the proportion of simulation iterates that converged to the posterior mode. We further assessed the mean run-time for each survextrap model fit.

## Appendix 2. Standard and smoothed M-spline basis functions

M-spline functions were originally designed to only model data within a pair of "boundary" knots [3]. However, parametric survival modelling, and extrapolation in particular, requires a hazard function $h(t)$ to be defined at all times $t$. Therefore, the survextrap model modifies the original M-spline specification [3] to assume a constant hazard after the final boundary. By default, for consistency with typical spline modelling practice, an additional smoothness constraint is defined by setting the derivative and second derivative of $h(t)$ at the upper boundary to be zero (see Appendix C from Jackson [4]). However, it is unclear whether this constraint is beneficial for modelling data in practice.

To address this, analyses were conducted where we evaluated models using the standard (without boundary constraint) and smoothed (with boundary constraint) set of M-spline basis functions, specified using the bsmooth argument. As the standard option requires at least 4 degrees of freedom, we only compared these models for df = 6 and 10.

When estimating 5-y RMST for Cetuximab OS using a survextrap model with df = 10 and $\sigma \sim$ Gamma(2,1), the estimation performance was comparable across both choices of M-spline basis, with relative bias 0.23% (0.14%) and 0.19% (0.14%) for a smoothed and standard basis, respectively. Applying these models to Nivolumab PFS, the relative bias using the standard basis was slightly higher at 0.80% (0.25%) compared to the default smoothed basis at -0.29% (0.24%) (Supplementary Figure 13). For both Cetuximab OS and Nivolumab PFS case studies, as we varied both the M-spline degrees of freedom and strength of the $\sigma$ smoothing prior, we found no material difference in model performance for estimating 5-y RMST.

**Supplementary Table 1:** Treatment effect functions for each of the three simulation scenarios.

| Scenario | Hazard ratios comparing the active relative to control arm |
|---|---|
| 1 | $HR(t) = 0.7$ |
| 2 | $HR(t) = -0.38 + 0.38 \tanh(0.8t - 1.2)$ <br><br> where $\tanh(x) = \frac{e^x + e^{-x}}{e^x - e^{-x}}$ is the hyperbolic tangent function. |
| 3 | $HR(t) = -2.8 \, f_{emg}(t; \mu = 0.8, \sigma = 0.4, \lambda = 0.35)$ <br><br> where $f_{emg}$ is the probability density function of an exponentially modified Gaussian distribution: <br><br> $f_{emg}(x; \mu, \sigma, \lambda) = \frac{\lambda}{2} e^{\frac{\lambda}{2}(2\mu + \lambda - 2x)} \text{erfc}\left(\frac{\mu + \lambda\sigma^2 - x}{\sqrt{2}\sigma}\right)$ <br><br> And $\text{erfc}$ is the complementary error function: <br><br> $\text{erfc}(x) = 1 - \text{erf}(x) = \frac{2}{\sqrt{\pi}} \int_x^\infty e^{-s^2} ds$ |

**Supplementary Table 2:** Stan model fitting mean run times for single arm case studies, comparing full MCMC with the Laplace optimisation 'opt' approximation.

| Case Study | Time (seconds) | Stan method | M-spline df |
|---|---|---|---|
| **Cetuximab OS** | 31.8 | MCMC | 3 |
| | 40.2 | MCMC | 6 |
| | 49.7 | MCMC | 10 |
| | 0.22 | Opt | 3 |
| | 0.30 | Opt | 6 |
| | 0.41 | Opt | 10 |
| **Nivolumab PFS** | 30.3 | MCMC | 3 |
| | 58.0 | MCMC | 6 |
| | 92.8 | MCMC | 10 |
| | 0.22 | Opt | 3 |
| | 0.30 | Opt | 6 |
| | 0.42 | Opt | 10 |

**Supplementary Table 3:** Stan model fitting mean run times for two-arm treatment effect scenarios, comparing full MCMC with the Laplace optimisation 'opt' approximation.

| Scenario | Time (seconds) | Stan method | Model | M-spline df |
|---|---|---|---|---|
| Proportional hazards | 110.2 | MCMC | PH | 10 |
| | 187.8 | MCMC | Non-PH | 3 |
| | 200.0 | MCMC | Non-PH | 6 |
| | 249.3 | MCMC | Non-PH | 10 |
| | 1.23 | Opt | PH | 10 |
| | 0.89 | Opt | Non-PH | 3 |
| | 1.21 | Opt | Non-PH | 6 |
| | 1.60 | Opt | Non-PH | 10 |
| Treatment waning | 105.7 | MCMC | PH | 10 |
| | 205.1 | MCMC | Non-PH | 3 |
| | 217.0 | MCMC | Non-PH | 6 |
| | 251.6 | MCMC | Non-PH | 10 |
| | 1.13 | Opt | PH | 10 |
| | 0.83 | Opt | Non-PH | 3 |
| | 1.07 | Opt | Non-PH | 6 |
| | 1.47 | Opt | Non-PH | 10 |
| Treatment delay then waning | 106.9 | MCMC | PH | 10 |
| | 164.3 | MCMC | Non-PH | 3 |
| | 199.6 | MCMC | Non-PH | 6 |
| | 249.7 | MCMC | Non-PH | 10 |
| | 1.33 | Opt | PH | 10 |
| | 0.88 | Opt | Non-PH | 3 |
| | 1.12 | Opt | Non-PH | 6 |
| | 1.56 | Opt | Non-PH | 10 |

**Supplementary Figure 1:** Survival and hazard plots for a survextrap model fitted using a random walk prior on the spline coefficients based on 50 simulated datasets from the Cetuximab OS case study.

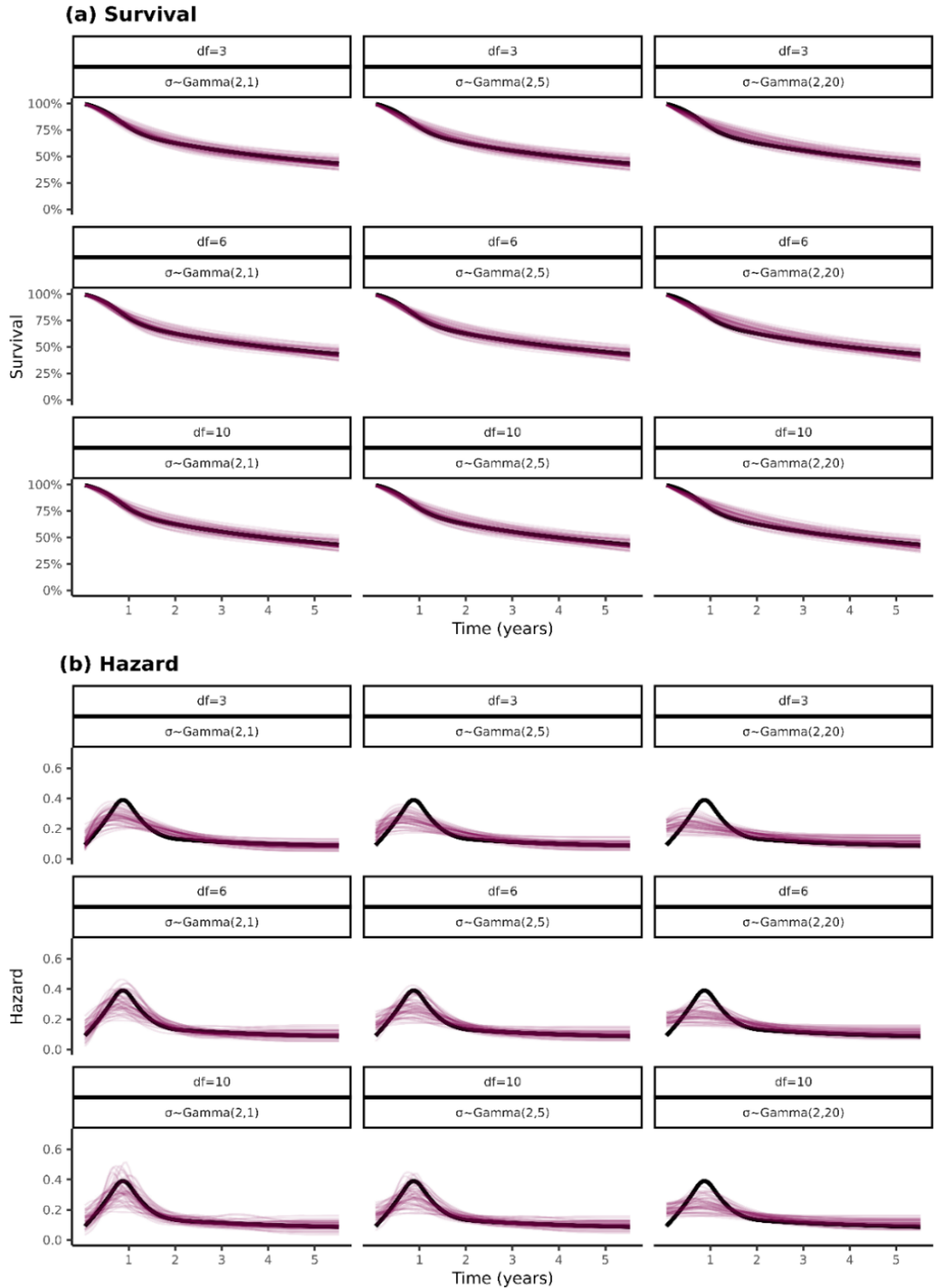

The black line shows the true survival and hazard function, the maroon lines show the model estimates.

**Supplementary Figure 2:** Survival and hazard plots for a survextrap model fitted using an exchangeable prior on the spline coefficients based on 50 simulated datasets from the Cetuximab OS case study.

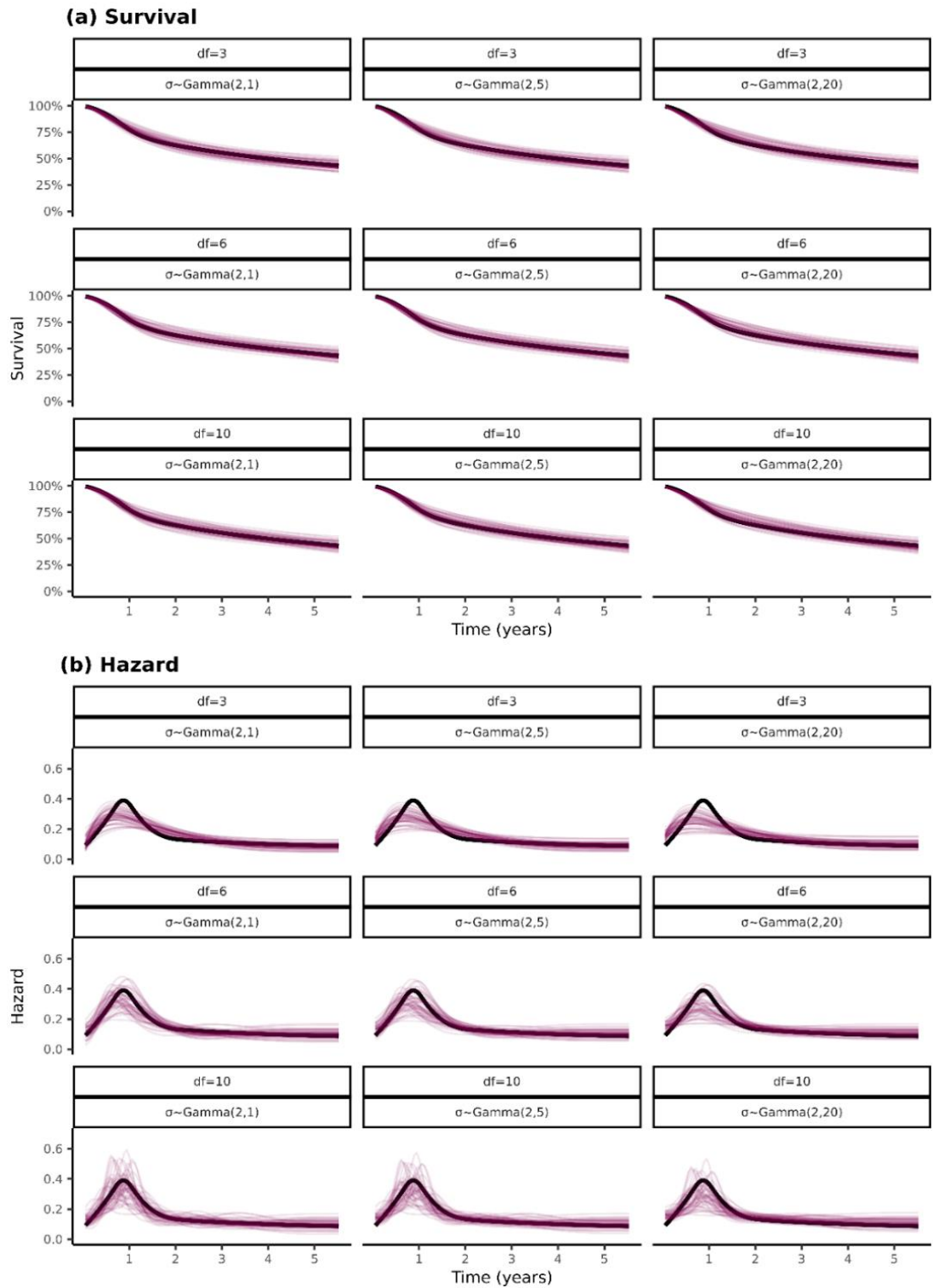

The black line shows the true survival and hazard function, the maroon lines show the model estimates.

**Supplementary Figure 3:** Performance measures for the posterior median RMST at 5-y based on simulated data from the Cetuximab OS and Nivolumab PFS case studies using an exchangeable prior for survextrap.

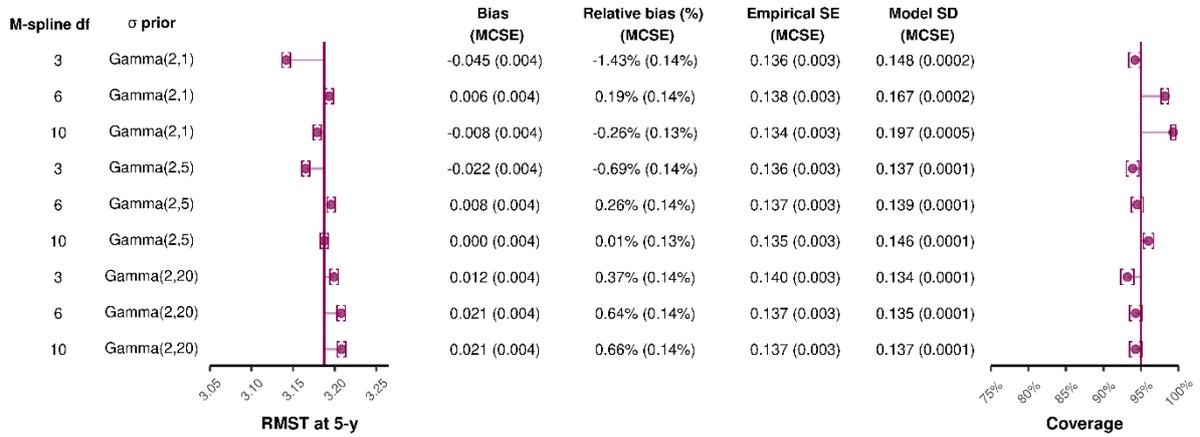

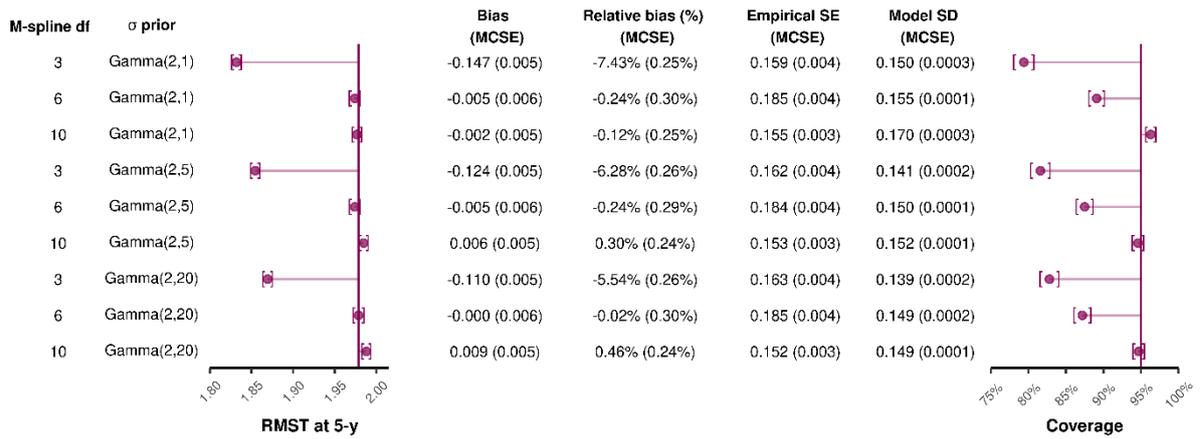

**Supplementary Figure 4:** Stan warning messages and convergence diagnostics for modelling a single treatment arm using a random walk prior on the spline coefficients.

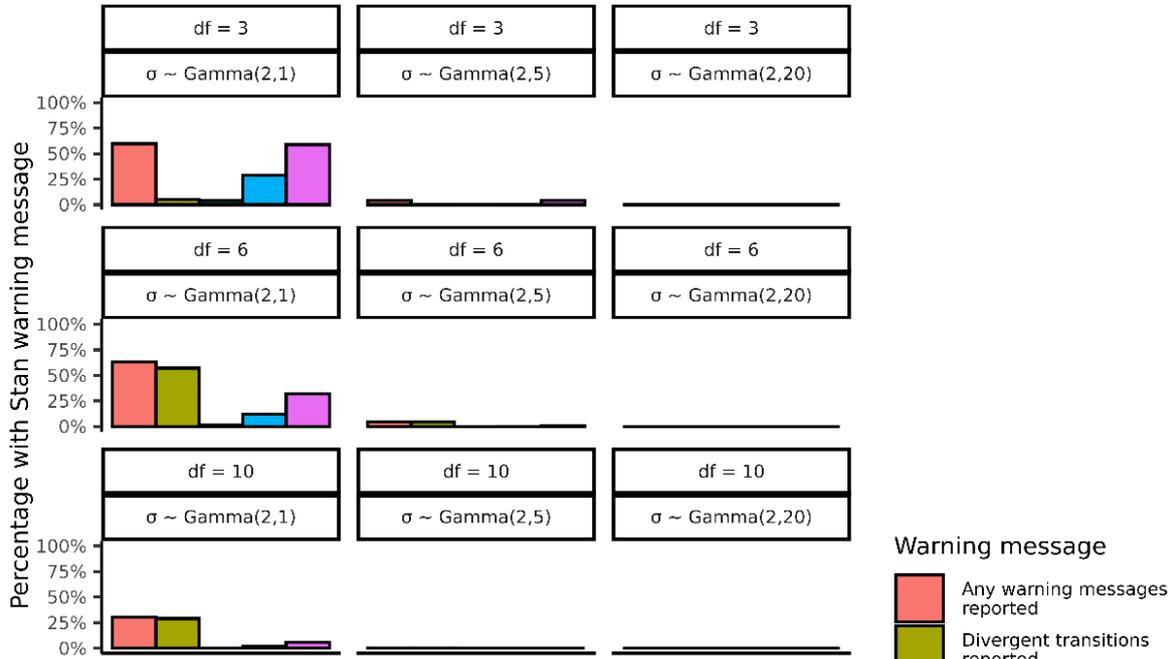

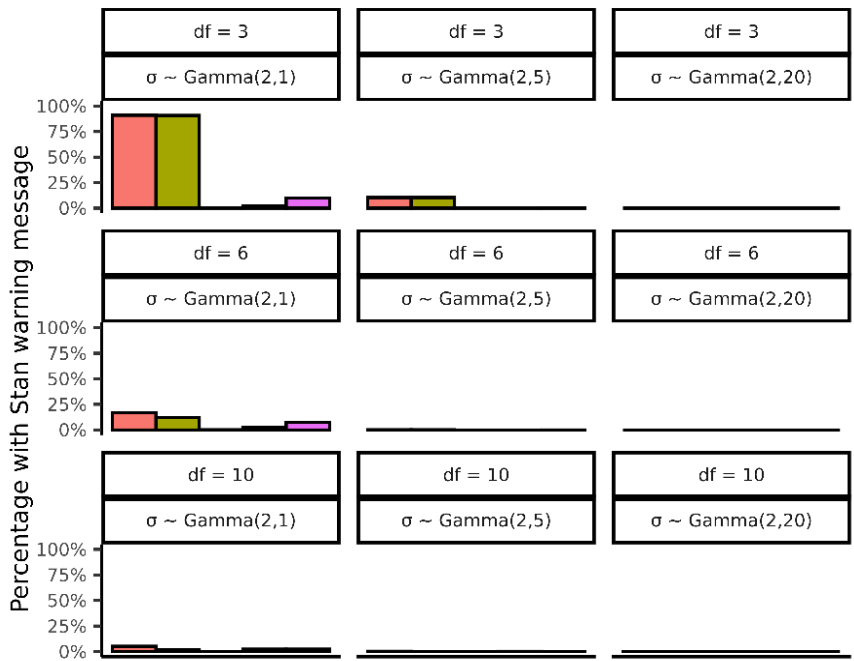

**Supplementary Figure 5:** Stan warning messages and convergence diagnostics for modelling a single treatment arm using an exchangeable prior on the spline coefficients.

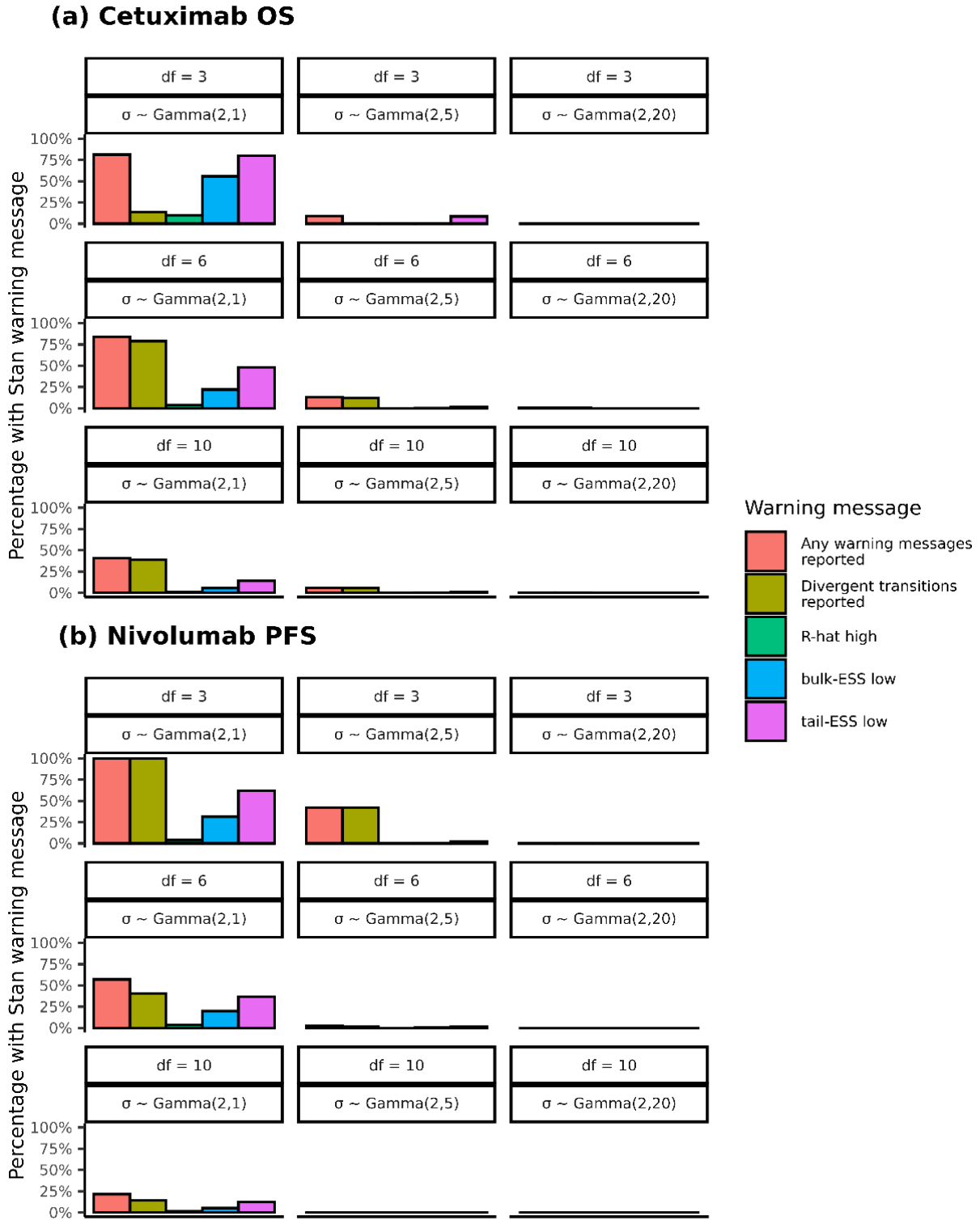

**Supplementary Figure 6:** Survival and hazard plots for a survextrap model fitted using a random walk prior on the spline coefficients based on 50 simulated datasets from the Nivolumab PFS case study.

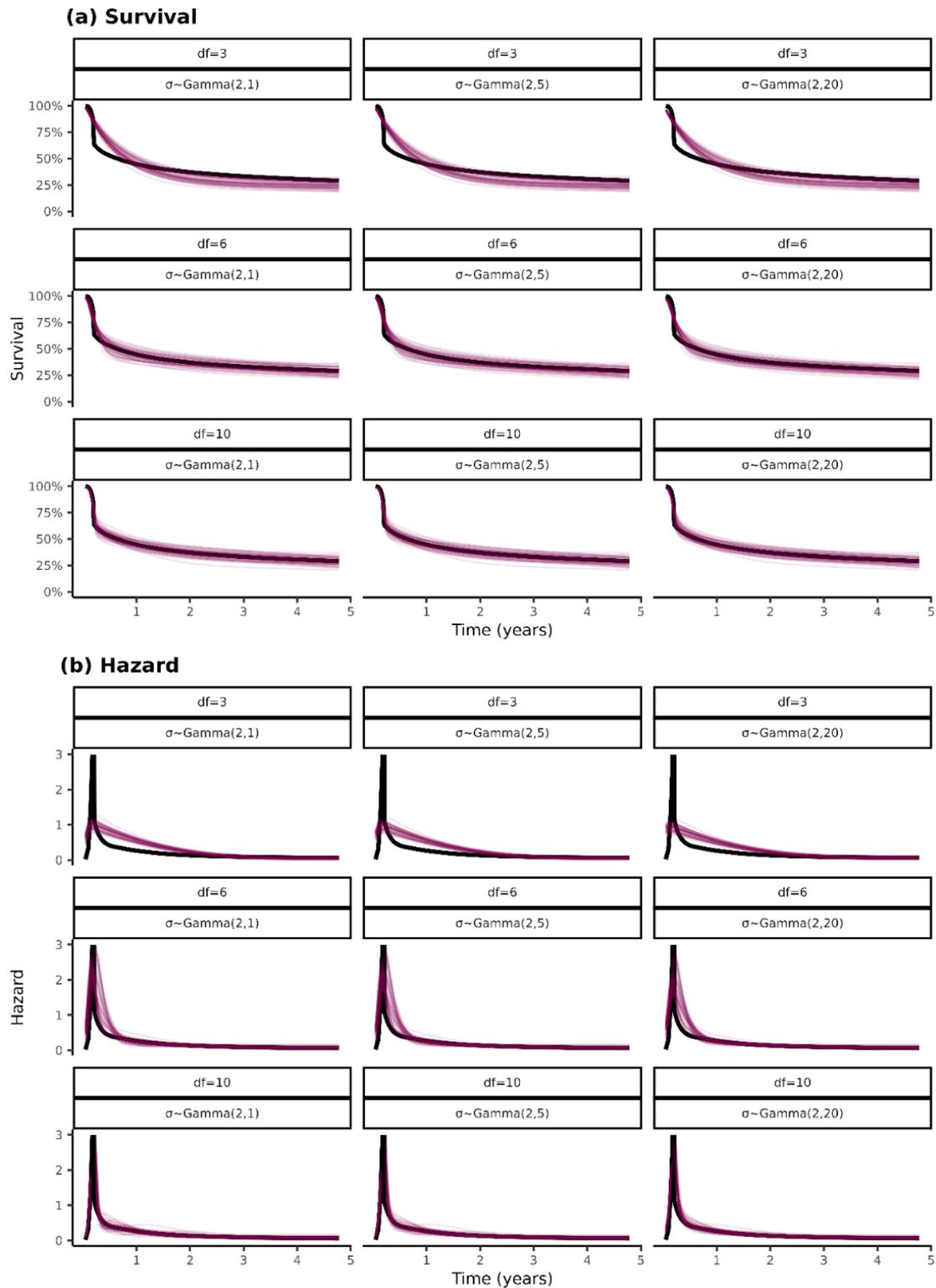

The black line shows the true survival and hazard function, the maroon lines show the model estimates.

**Supplementary Figure 7:** Survival and hazard plots for a survextrap model fitted using an exchangeable prior on the spline coefficients based on 50 simulated datasets from the Nivolumab PFS case study.

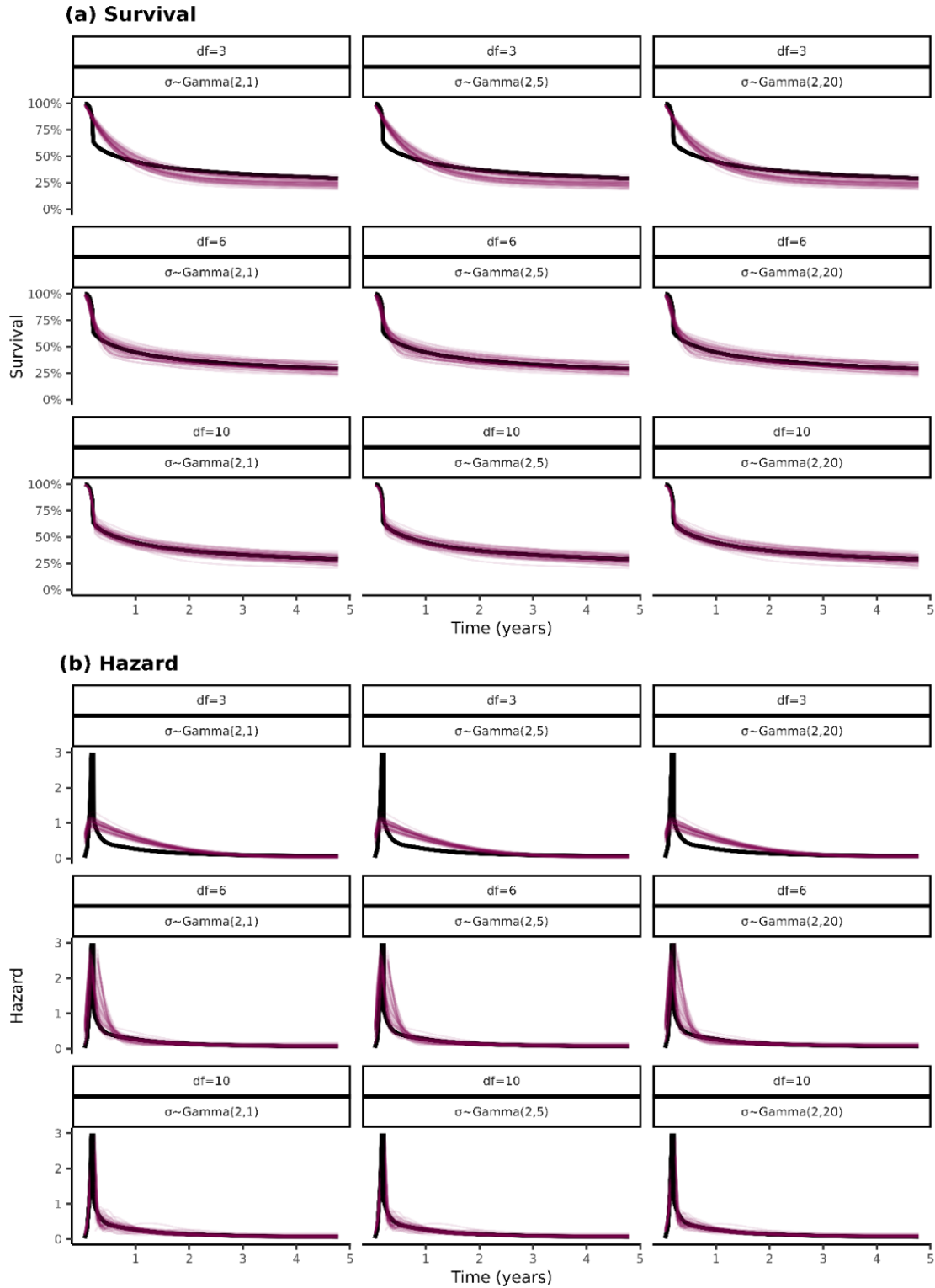

The black line shows the true survival and hazard function, the maroon lines show the model estimates.

**Supplementary Figure 8:** Performance measures for the posterior median RMST at 5-y based on simulated data from the Cetuximab OS and Nivolumab PFS case studies using a Laplace approximation method (fit_method = "opt") for survextrap and a random walk prior.

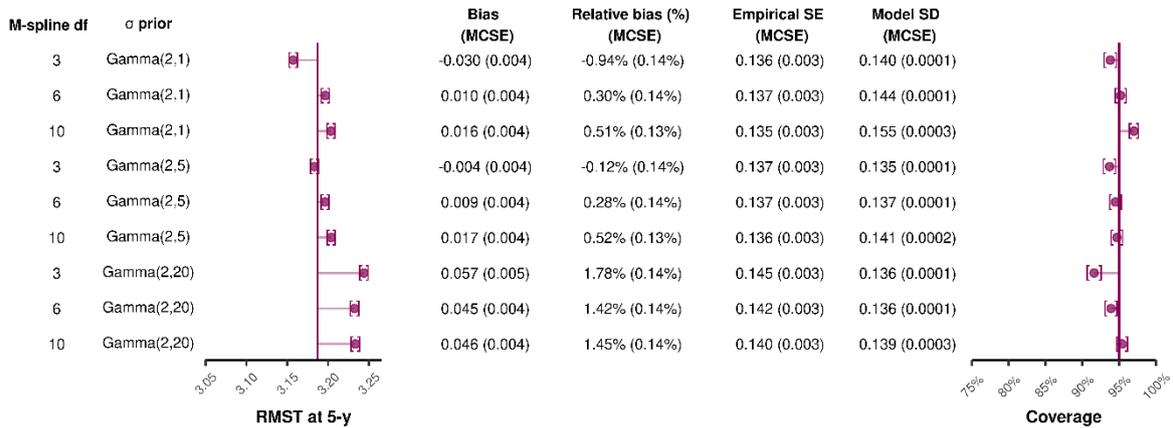

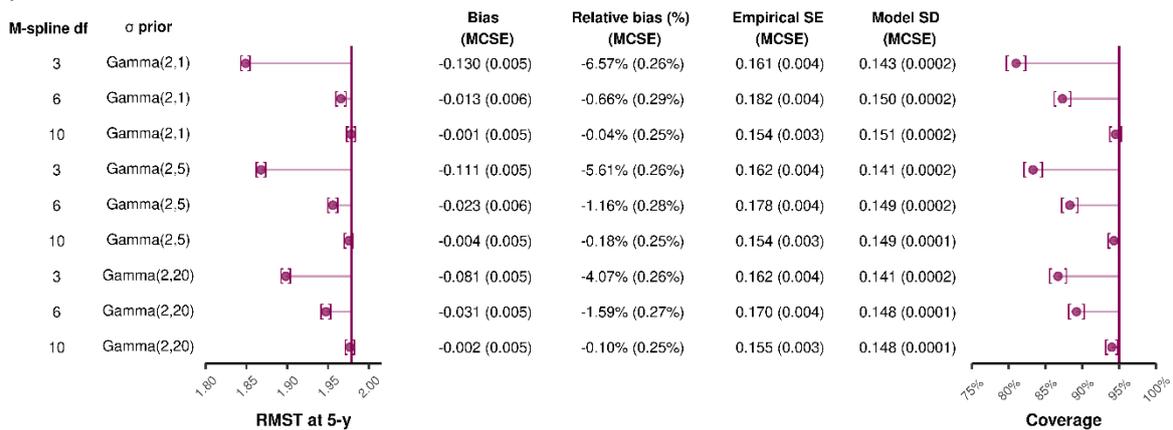

**Supplementary Figure 9:** Performance measures for the RMST at 5-y based on simulated data from the Cetuximab OS and Nivolumab PFS case studies using frequentist models implemented in R packages flexsurv and rstpm2.

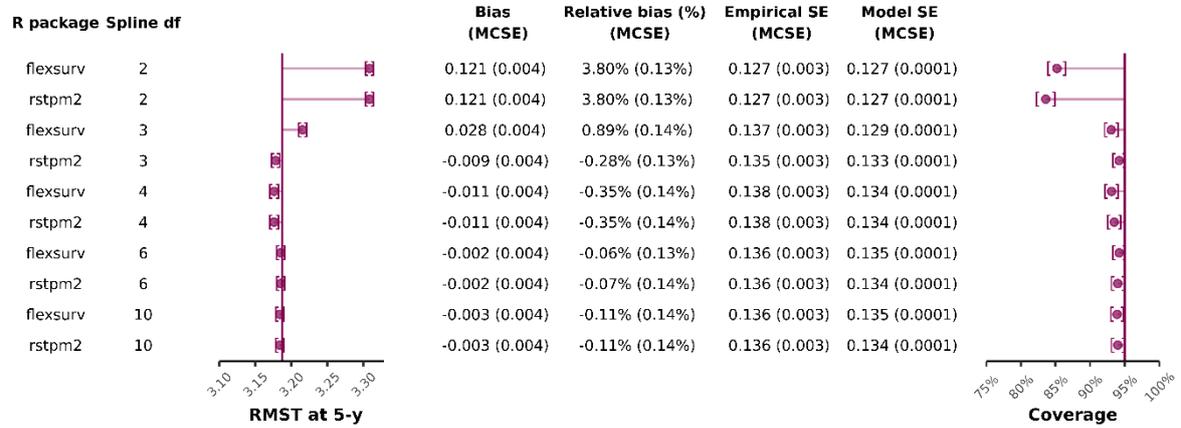

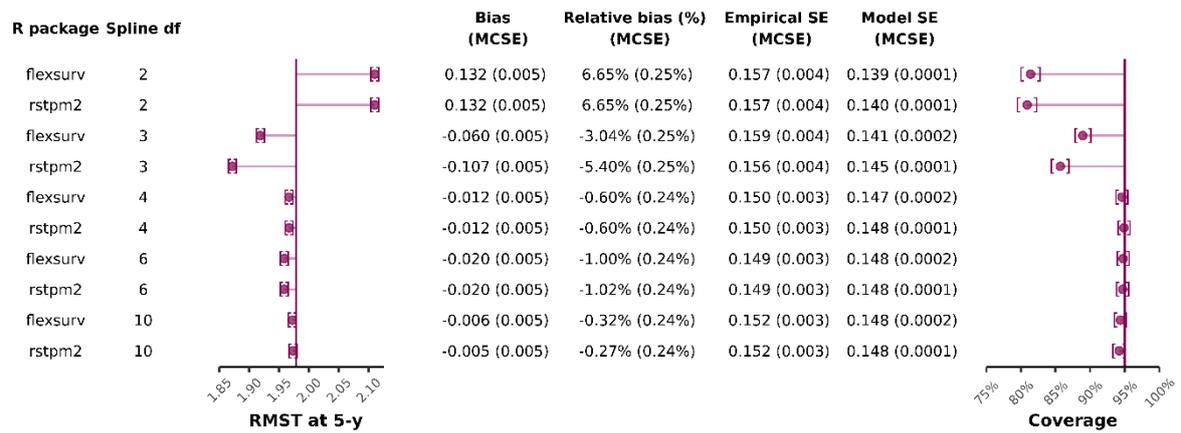

**Supplementary Figure 10:** Performance measures for the difference in RMST at 5-y based on simulated data from the Cetuximab OS case study for the control arm and under three scenarios for the time-varying treatment effect, investigating fitted survextrap models that use an exchangeable prior and a default Gamma(2,1) prior for $\sigma$ and where the degrees of freedom and prior for $\tau$ are varied.

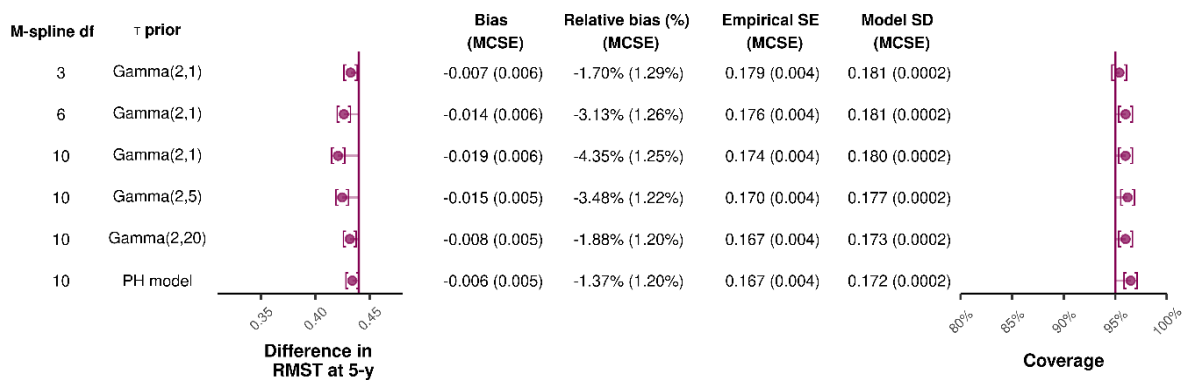

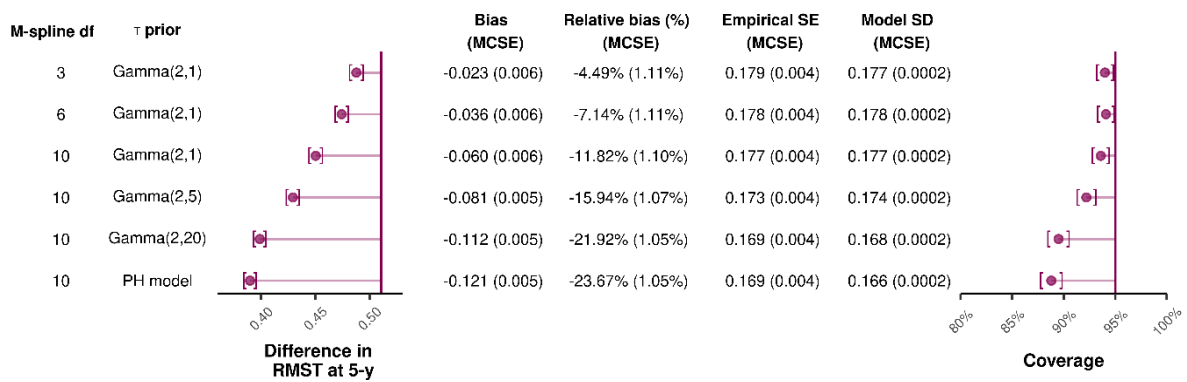

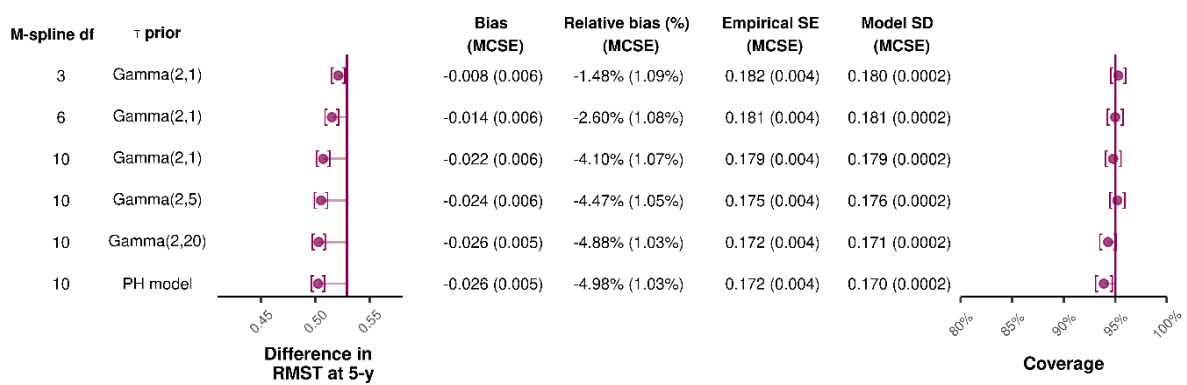

**Supplementary Figure 11:** Hazard ratio plots for a non-proportional hazards survextrap model fitted using an exchangeable prior and varying the degrees of freedom and non-proportionality smoothness prior $\tau$, based on 50 simulated datasets under three treatment-effect scenarios.

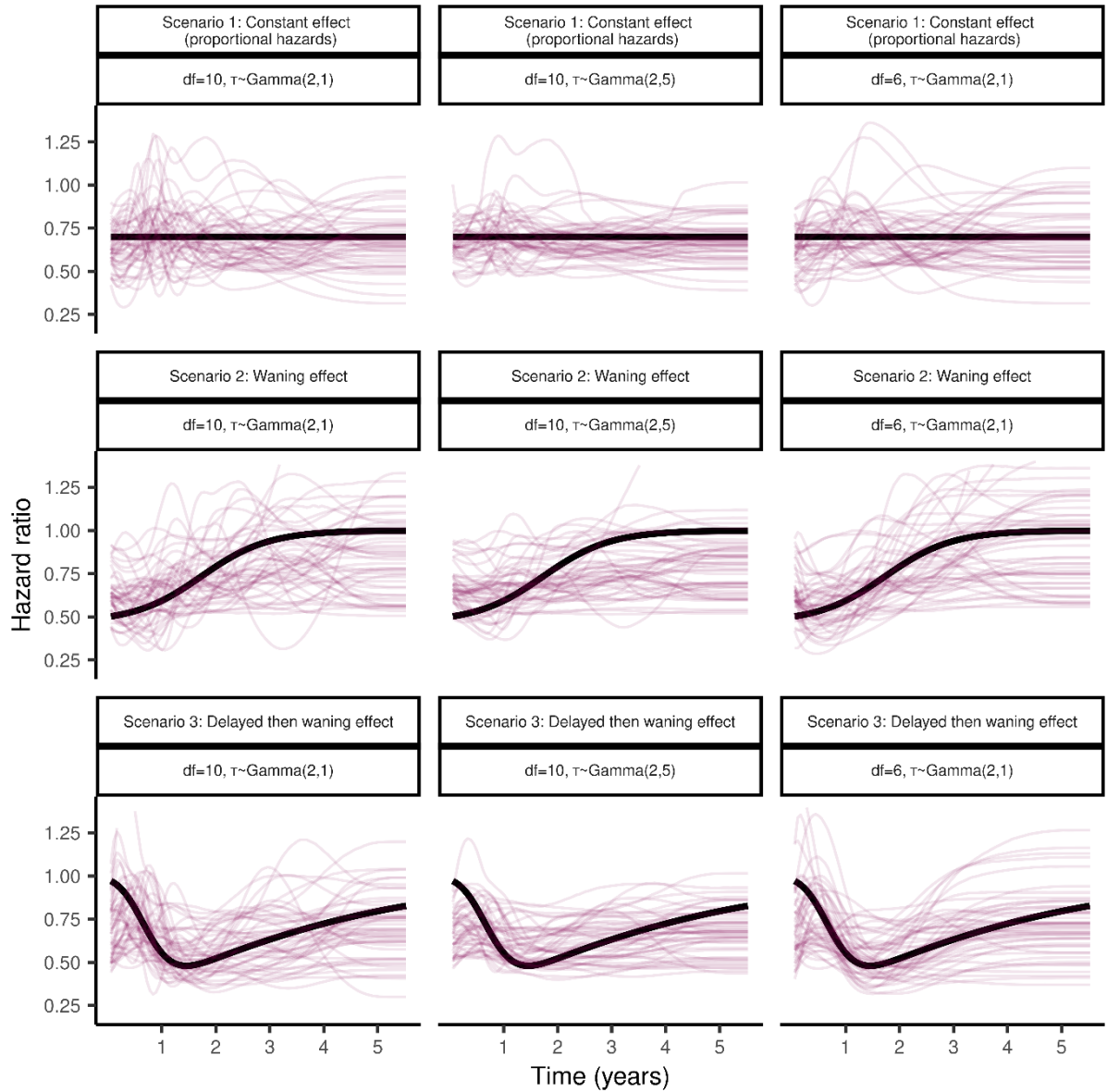

**Supplementary Figure 12:** Performance measures of frequentist methods for estimating the difference in RMST at 5-y.

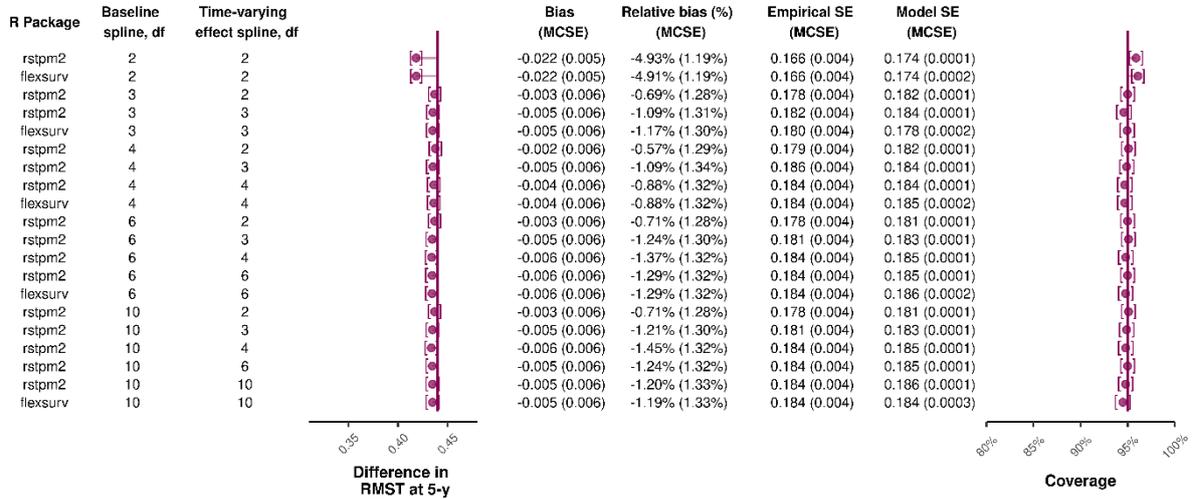

(a) Scenario 1: Constant effect (proportional hazards)

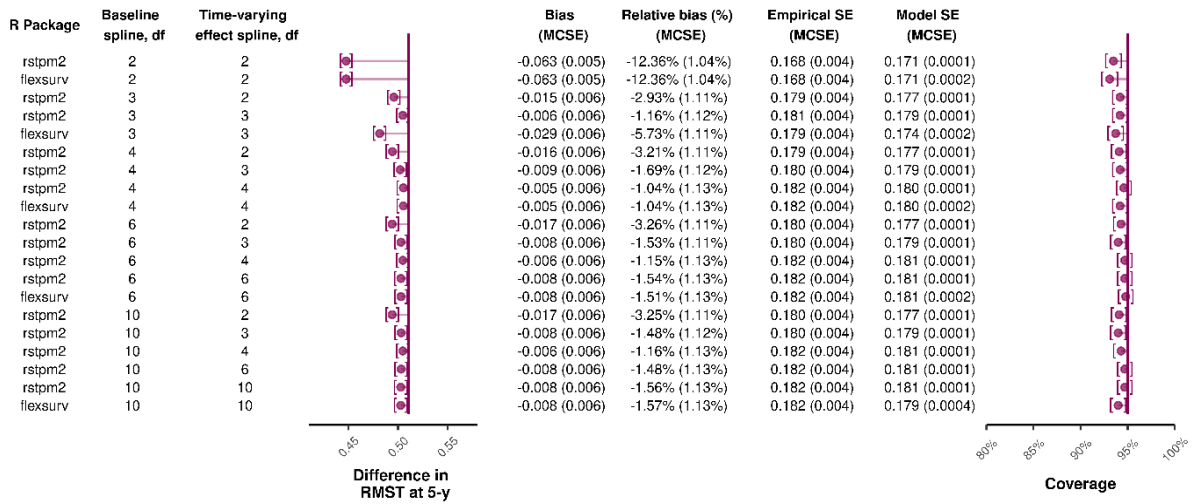

(b) Scenario 2: Waning effect

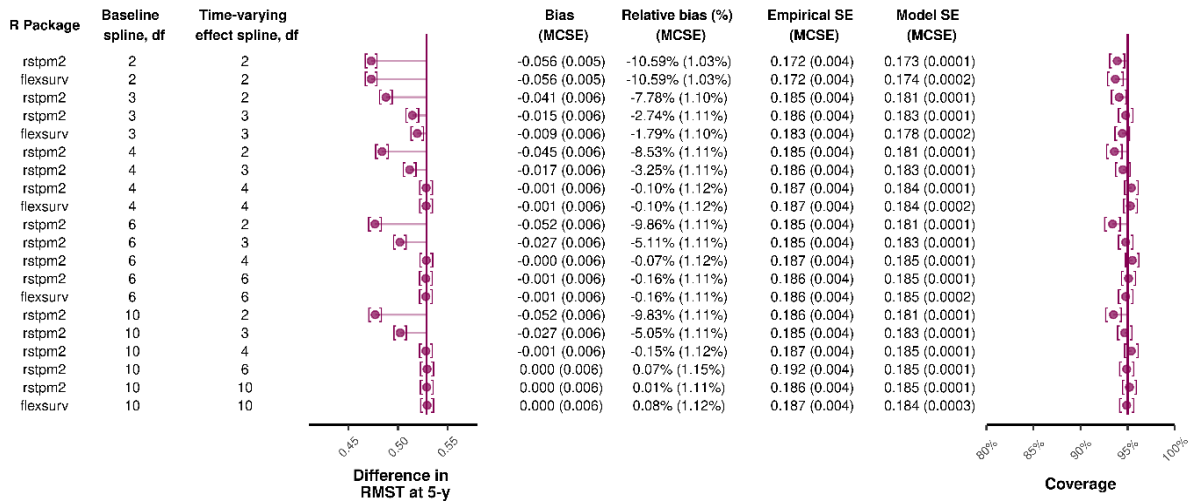

(c) Scenario 3: Delayed then waning effect

**Supplementary Figure 13:** Performance measures for estimating 5-y RMST using the standard M-spline basis functions (unsmoothed at the upper boundary knot, option bsmooth = FALSE) and smoothed basis functions (smoothed at upper boundary knot, option bsmooth = TRUE). Models fitted using a random walk prior.

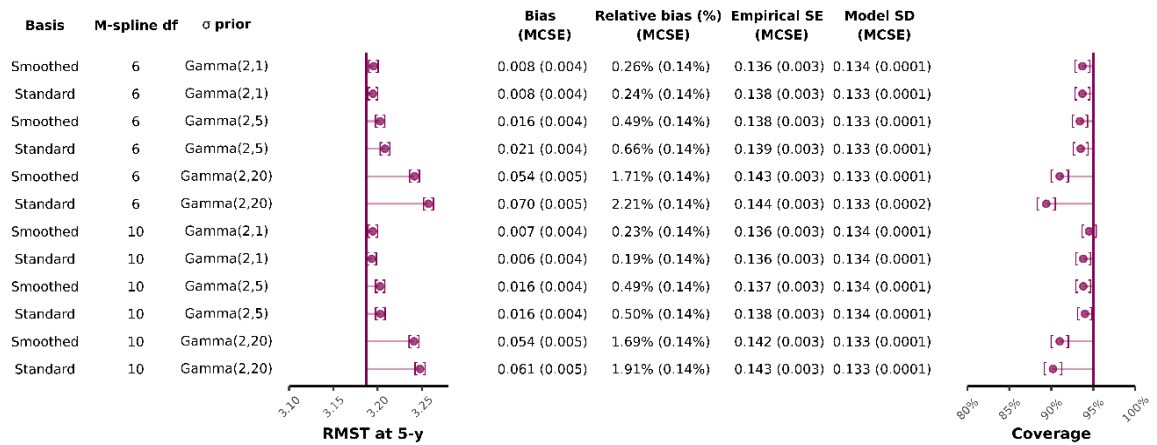

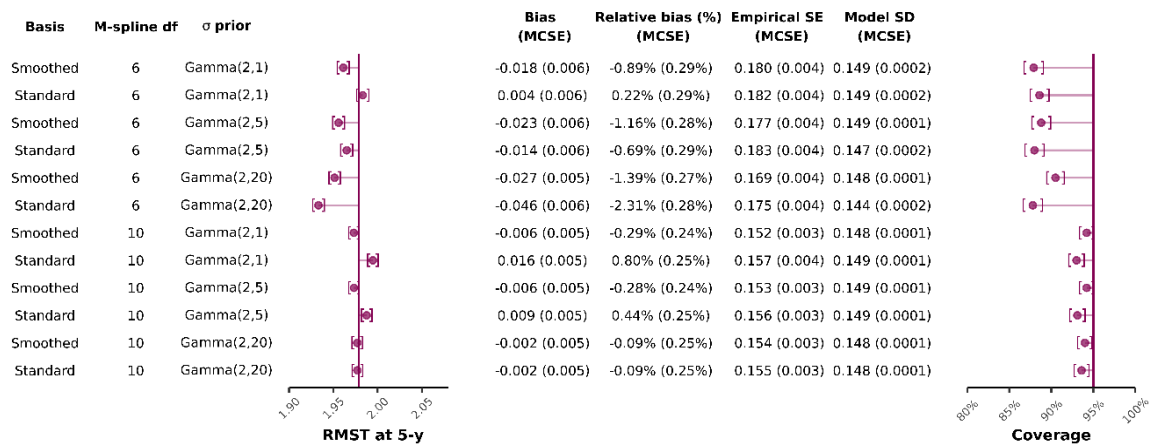

# Supplementary Information - References